\documentclass[preprint,aps]{revtex4} 
\usepackage[latin9]{luainputenc}
\usepackage{float}
\usepackage{amsmath}
\usepackage{amssymb}
\usepackage{graphicx}
\usepackage{setspace}

\makeatletter

\@ifundefined{textcolor}{}
{%
 \definecolor{BLACK}{gray}{0}
 \definecolor{WHITE}{gray}{1}
 \definecolor{RED}{rgb}{1,0,0}
 \definecolor{GREEN}{rgb}{0,1,0}
 \definecolor{BLUE}{rgb}{0,0,1}
 \definecolor{CYAN}{cmyk}{1,0,0,0}
 \definecolor{MAGENTA}{cmyk}{0,1,0,0}
 \definecolor{YELLOW}{cmyk}{0,0,1,0}
}

\usepackage{amsfonts}
\usepackage{bm}
\usepackage[dvips]{epsfig}
\usepackage[dvips]{graphicx}
\usepackage{float}
\usepackage{dcolumn}
\usepackage{ifpdf}
\usepackage{multirow}

\newcommand{\be}{\begin{equation}}
\newcommand{\ee}{\end{equation}}
\newcommand{\bes}{\begin{subequations}}
\newcommand{\ees}{\end{subequations}}
\newcommand{\ben}{\begin{eqnarray}}
\newcommand{\een}{\end{eqnarray}}

\makeatother

\begin{document}

\title{Scattering of metastable lumps in a model with a false vacuum}

\author{Fred C. Lima$^{1}$, Fabiano C. Simas$^{1,2}$, K. Z. Nobrega$^{3}$, Adalto R. Gomes$^{1}$
}
\email{fredfjcl@gmail.com,fc.simas@ufma.br,bzuza1@yahoo.com.br,argomes.ufma@gmail.com}


\affiliation{
$^{1}$ Programa de P\'os-Gradua\c c\~ao em F\'isica, Universidade Federal do Maranh\~ao
(UFMA), Campus Universit\'ario do Bacanga, 65085-580, S\~ao Lu\'is, Maranh\~ao, Brazil \\
 $^{2}$ Centro de Ci\^encias Agr\'arias e Ambientais-CCAA, Universidade
Federal do Maranh\~ao (UFMA), 65500-000, Chapadinha, Maranh\~ao, Brazil\\
$^{3}$ Departamento de Engenharia Teleinform\'atica, Universidade Federal do Cear\'a (UFC),
60455-640, Fortaleza, Cear\'a, Brazil
 }
\begin{abstract}

In this work we consider the scalar field model with a false vacuum proposed by A. T. Avelar, D. Bazeia, L. Losano and R. Menezes, Eur. Phys. J. C 55, 133-143 (2008). The model depends on a parameter $s>0$.  The model has unstable nontopological lump solutions with a bell shape for small $s$, acquiring a flat plateau around the maximum for large $s$. For $s\to\infty$ the $\phi^4$ model is recovered. We show that for 
$s\gtrsim 2$ the lump is metastable with the only negative mode very close to zero. Metastable lumps can propagate and survive long enough to produce dynamical effects. Due to their simplicity, they can
be an alternative to the procedure of stabilization which requires, for instance, a complex scalar field to construct nontopological solitons. We study lump-lump collisions in this model, describing the main characteristics of the scattering products at their dependence on $s$ and the initial velocity modulus of each lump.

\end{abstract}

\pacs{XXX}

\maketitle


\section{ Introduction }


In classical field theory, some solutions can be classified with respect to the topological  mapping between the space of coordinates and the internal space of configurations, and are of interest in physics \cite{raja,vile, dau}. In a scalar field theory in $(1,1)$ dimensions, the topological defect is the  kink (and antikink). In this case, the scalar field  $\phi(x)$ has different asymptotic limits for $x\to\pm\infty$ and the solution is linearly stable. The static kink solution $\phi_K(x)$ is related to the static antikink solution $\phi_{\bar K}(x)$ by $\phi_{\bar K}(x)=-\phi_{K}(x)$. The nontopological defect is named lump \cite{wile, sut}. In this case, the scalar field has equal asymptotic limits for $x\to\pm\infty$ and is unstable under linear perturbations. 

Kink-like defects are subject of many investigations in several areas of physics. In particular, these structures can appear in cosmology \cite{giblin,agui}, optics communications \cite{agraw,molle}, superfluids \cite{yasui}, ferroelectrics \cite{struk} and DNA \cite{yakus}. Lump-like defects are also of interest describing,  for instance, bright solitons in optical fibers \cite{agra, haus} and localized excitations in Bose-Einstein condensates \cite{ab}. 
In $(3,1)$ dimensions, nontopological scalar field solutions were investigated in early universe cosmology \cite{cosm1,cosm2,cosm3,cosm4}, and considered as a model for dark matter \cite{ponton}.  In some of these applications, the lump can be stabilized if described by a complex scalar field, as done in q-ball models \cite{qb1}, or coupled with other charged matter fields \cite{baz1}. In this case the stable solution has a  conserved Noether charge and is named by some authors as nontopological solitons. 

Models with analytic static lump solutions were studied in the Refs. \cite{baz1,baz2,avelar1,avelar2}. In the Ref. \cite{campos1}, the authors investigated the fermion transfer due to the scattering between an antikink and a lump in the $\phi^4$ model with a Yukawa coupling.  The collisions between lump and solitary waves in $(3,1)$ dimensions were studied in the context of extended Kadomtsev-Petviashvili equation, which describes the multi-component plasma model \cite{liu2}. In the Ref. \cite{kurt}, the generation and the evolution of lump Benney-Luke solitary waves was analyzed. These are solutions of the Kadomtsev-Petviashvili I equation, which model small amplitude shallow-water waves. The existence of such waves for the Benney-Luke equation with surface tension it has been discussed in the Ref. \cite{pego}, and in appropriate limits, this equation reduces to the integrable Korteweg-de Vries (KdV) equation. In the Ref. \cite{safi} it was discussed the scattering between lumps with periodic waves, with other lumps and with kink solutions. There, it was observed interesting effects such as the fission of lump waves. Motivated by the interest in non-diffractive and non-dispersive wave packets propagating in optical media, in the Ref. \cite{fabio} the authors showed the existence and interactions of dark-lump solitary wave solutions of the  nonlinear Schr\"odinger equation in $(2,1)$ dimensions. The results showed connections between nonlinear wave propagation in optics and hydrodynamics. 

In this work, we considered the lump scattering in $(1,1)$ dimensions in a model with false vacuum that is an extension of the $\phi^4$ model. In the next section, we introduce the model and the nontopological solutions. Furthermore, we also discuss the linear stability analysis of the solutions. In the Sect. III, we present the structure of lump-lump scattering. We report the formation of localized oscillations and kink-antikink pairs. We present our main conclusion in the Sect. IV.


\section{The Model}


We consider the following action with standard dynamics (we work in a flat metric with signature  $(+-)$)
\begin{eqnarray}
S=\int{dxdt\bigg(\frac{1}{2}\partial_{\mu}\phi\partial^{\mu}\phi-V(\phi)\bigg)}.
\end{eqnarray}
The equation of motion is given by
\begin{equation}\label{eom}
	\frac{\partial^2 \phi}{\partial t^2} - \frac{\partial^2 \phi}{\partial x^2} + \frac{dV(\phi)}{d\phi}=0.
\end{equation}
Static solutions that minimize the energy satisfy the following first-order equations: 
\begin{equation}\label{first_order_eq}
	\frac{d\phi}{dx} = \pm \sqrt{2V(\phi)}.
\end{equation}
For topological (kink-like) solutions, each equation in (\ref{first_order_eq}) provides a solution in $-\infty<x<\infty$. On the other hand, for the nontopological lump solutions studied here we have the following conditions \cite{avelar1,avelar2}
\begin{equation}\label{eq_1a_order_lump}
\frac{d\phi}{dx}= \sqrt{2V(\phi)} \text{ for } x<0 \hspace{0.5 cm} \text{and} \hspace{0.5 cm } \frac{d\phi}{dx}= - \sqrt{2V(\phi)} \text{ for } x>0.
\end{equation}

\begin{figure}
	\includegraphics[width=8cm]{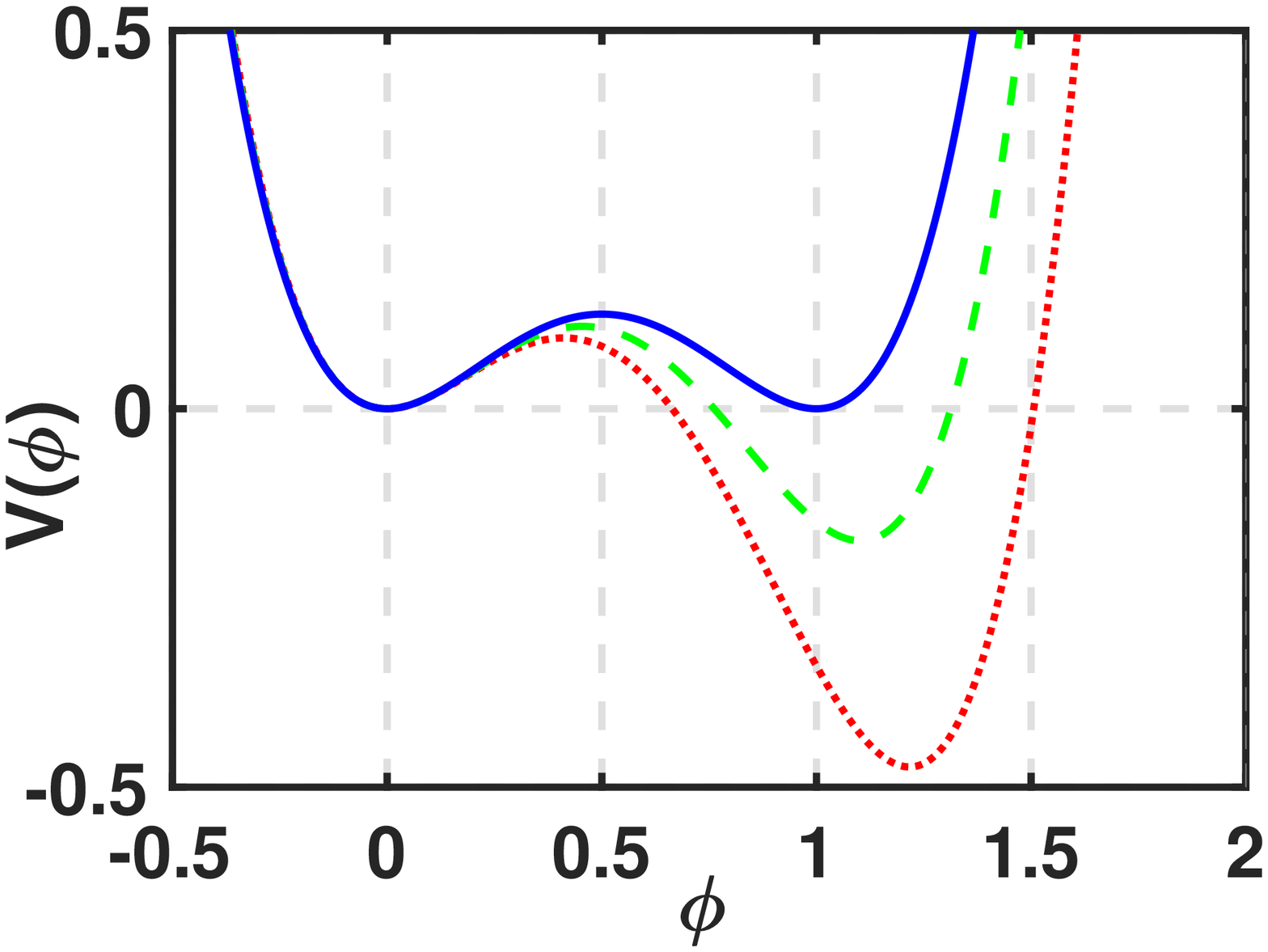} 
	\includegraphics[width=8cm]{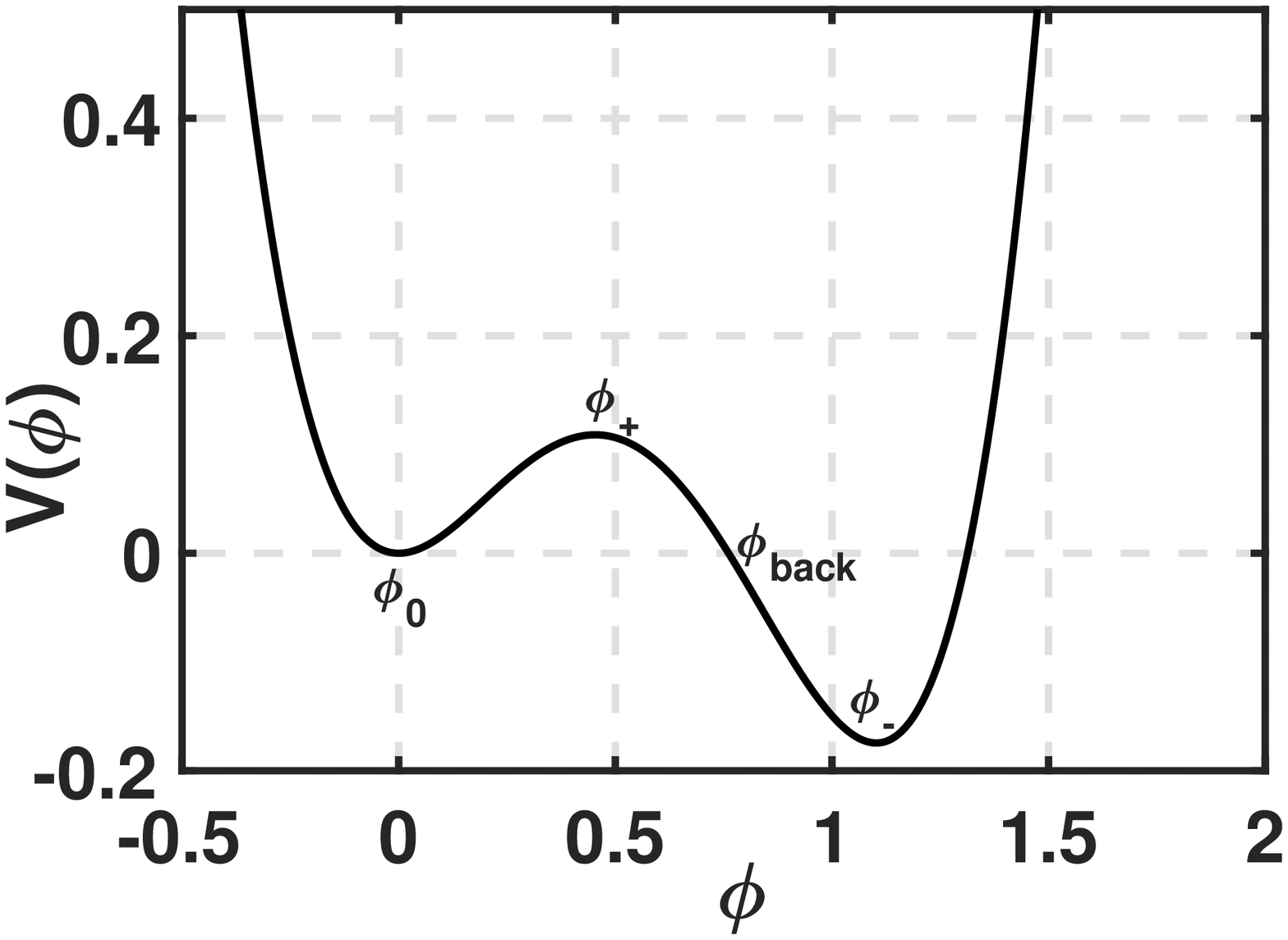}
	\caption{a) Potential $V(\phi)$ for $s=0.8$ (dotted red), $s=1$ (dash green) and $s=5$ (solid blue); b) Potential $V(\phi)$ for $s=1$, identifying the points $\phi_+$, $\phi_-$, $\phi_0$ e $\phi_{back}$.}
	\label{potential}
\end{figure}

 In this work, we consider the potential given by \cite{avelar2}
\begin{eqnarray}
V(\phi) =2\phi^2 \left( \phi - \tanh(s) \right)\left( \phi - \coth(s) \right),
\label{potential_eq}
\end{eqnarray}
where $ s>0 $ is a real parameter. In the limit $s \rightarrow \infty$ the $\phi^4$ model is achieved.  The Fig. \ref{potential}a shows this potential for some value of s. The Fig. \ref{potential}b shows the main characteristics of the potential for $s=1$. For general $s$, the potential has a local minimum at 
\be 
\phi_0=0,
\ee
a global minimum at 
\be 
\phi_- = \frac34 \coth(2s) + \frac14 \sqrt{9\coth^2(2s) -8},
\ee
 and a local maximum at 
\be 
\phi_+ = \frac34 \coth(2s) - \frac14 \sqrt{9\coth^2(2s) -8}
\ee
and crosses the $\phi$ axis at
\be 
\phi_{back} = \tanh(s).
\ee
 We note that the value of $V(\phi)$ at $\phi_+$ is always positive. On the other hand, the value of the potential at $\phi_-$ is negative, but getting to zero for $s \rightarrow \infty$. Then, the model is characterized by a false vacuum at $\phi_0$ and a true vacuum at $\phi_-$.

\begin{figure}
	\includegraphics[width=10cm]{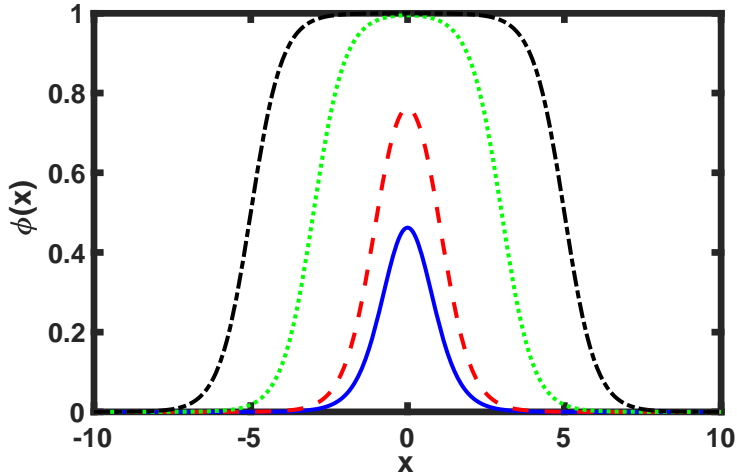} 
	\caption{Lump solution $\phi(x)$ for $s=0.50$ (solid blue), $s=1$ (dash red), $s=3$ (dotted green) and $s=5$ (dotted-dash black).}
	\label{fig_sol_static}
\end{figure}

The lump solution of the Eq. (\ref{eq_1a_order_lump}) centered at $x=0$ is given by \cite{avelar2}
\begin{eqnarray}
\phi_s(x)=\frac{1}{2} \left(\tanh(x+s) - \tanh(x-s)\right).
\label{eq:sol_static}
\end{eqnarray}
The Fig. \ref{fig_sol_static} shows the profile of the static solution for some values of s.  Comparing the Figs. \ref{potential}b and \ref{fig_sol_static} we note that the lump solution starts at the local minimum $\phi=\phi_0$ of the potential, increases until $\phi=\phi_{back}$ and return to the local minimum.  The parameter $s$ controls the width of the lump-like solutions. Small $s$ corresponds to bell-shaped lump. However, for large values of $s$ the static solution has a flat plateau around the maximum. This is an unusual structure for lumps and a good reason for studying such solutions.

\begin{figure}
	\includegraphics[width=10cm]{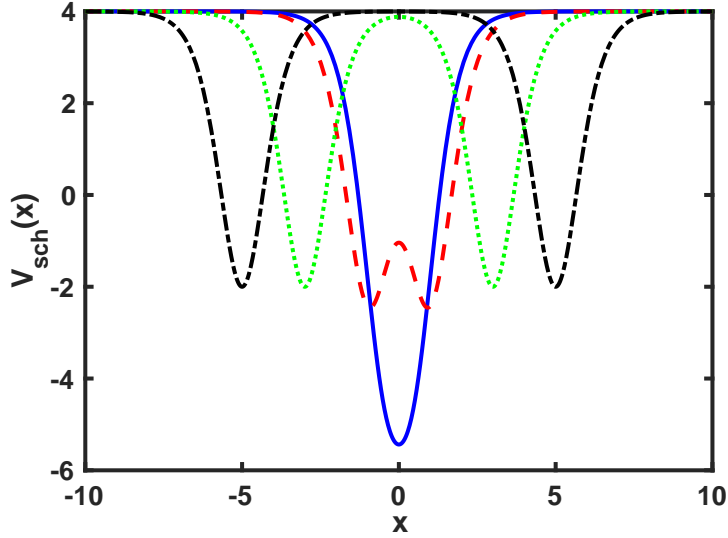} 
	\caption{Schr\"odinger-like potential for $s=0.50$ (solid blue), $s=1$ (dash red), $s=3$ (dotted green) and $s=5$ (dotted-dash black). }
	\label{potential_vsch}
\end{figure}

The linear stability analysis considers perturbations around the static solution of the form $ \phi(x,t)=\phi_s (x)+\eta (x) \cos(\omega t)$. This leads to a Schr\"odinger-like equation,
\begin{eqnarray}\label{eq_de_sch}
-\frac{d^2 \eta}{dx^2} + V_{sch}(x)\eta(x)=\omega^{2}\eta,
\end{eqnarray}
with $ V_{sch}(x)=V_{\phi\phi}(\phi) $ at $\phi=\phi_s (x)$. Explicitly, this gives
\begin{eqnarray}
V_{sch}(x)=4+24\phi_{s}^{2}(x) - 12\left[\tanh(s) + \coth(s)\right]\phi_{s}(x).
 \label{Vsch}
\end{eqnarray}

A plot for this potential for several values of $s$ is depicted in the Fig. \ref{potential_vsch}. This potential can be separated in two regions from the value $ \bar{s} = (1/2)\ln(\sqrt{3} + 2)\simeq0.66 $ \cite{bazeia1}. From $ s\le \bar{s} $ the potential has a single-well, whereas for $ s >\bar{s} $ the potential takes the shape of a double-well potential. Moreover, for $s\gtrsim3$ the potential acquires a plateau around $x=0$ whose tickness increase with $s$ \cite{avelar2}. In this region the potential is similar to that of a kink-antikink pair.

\begin{figure}
	\includegraphics[width=10cm]{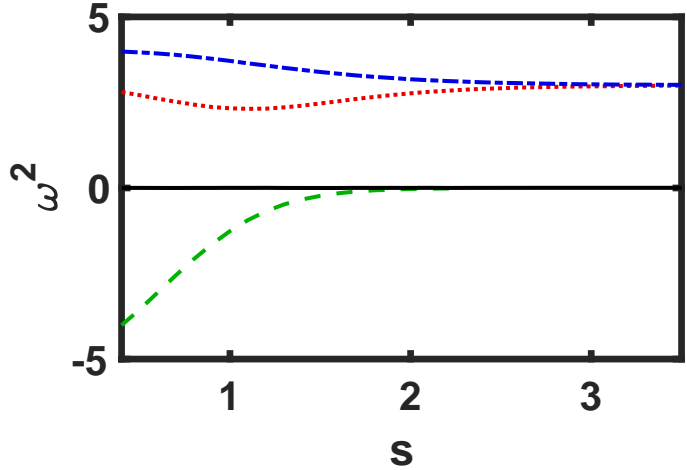} 
	\caption{Eigenvalues $\omega^2$ of the Schr\"odinger-like equation as a function of $s$. One can see the negative mode (dash green), the translational mode (solid black) and the first (dotted red) and the second (dotted-dash blue) vibrational mode.}
	\label{spectrum_freq}
\end{figure}

We solved the Schr\"odinger-like Eq. (\ref{eq_de_sch}) and plotted the results of $\omega^2$ as a function of $s$ in the Fig. \ref{spectrum_freq}. The figure shows for low values of $s$ the presence of four discrete modes: i) a negative energy mode, characterizing instability, ii) a zero-mode or translational mode, iii) two vibrational modes, with $\omega^2>0$. The plot shows a monothonic increasing of the energy of the unstable mode, with $\omega^2$ still negative, but very close to zero for $s\gtrsim2$. This shows that the lump solution in this region is metastable, living sufficiently in some cases to produce dynamical effects. The two vibrational modes of the Fig. \ref{spectrum_freq} with positive energy coalesce to only one vibrational mode for $s\gtrsim 3$, where the spectrum is already very close to that of a kink-antikink pair in $\phi^4$. Then, for $0<s\lesssim2$ one has unstable lumps, for $2\lesssim s \lesssim 3$ a long-lived lump state with multiple vibrational states, whereas for $s\gtrsim 3$  a long-lived lump with one vibrational state.

\begin{figure}
	\includegraphics[width=12cm]{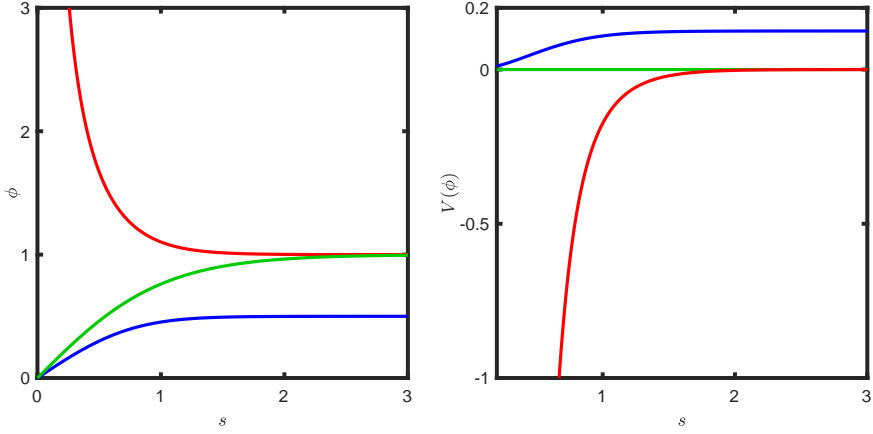}
	\caption{ a) $\phi_-$ (red), $\phi_{back}$ (green) and $\phi_+$ (blue)  and b) corresponding potential  $V(\phi)$ as a function of $s$. The lump solution interpolates between $\phi_0=0$ and $\phi_{back}$. In both points the potential $V$ is zero.}
	\label{phi_V_s}
\end{figure}

The metastable character of the lump solution for large $s$ is also evident in the Figs. \ref{phi_V_s}a-b. In the Fig. \ref{phi_V_s}a one can see the dependence of the global minimum $\phi_-$, the local maximum $\phi_+$ and $\phi_{back}$ as a function of $s$. In the Figs. \ref{phi_V_s}b one can see the corresponding values of the potential in these minima. Note that, for $s\gtrsim 2$ the negative global minima is too close from $V=0$, the value of the scalar field corresponding to $\phi_{back}$ and to $\phi_0=0$. This means that for these values of $s$ the lump is a bell-shaped with a plateau, with the scalar field at $x\to\pm\infty$ corresponding to the potential very close to the global minimum (the true vacuum $\phi_-$). The same applies to the value $\phi_{back}$, the value of the scalar field at the center of the lump. This is the point where the potential $V(\phi)$ crosses the axis $\phi$ (see the Fig. \ref{potential}), with $V(\phi_0)=0$. This, for $s\gtrsim 2$ is also very close to the true vacuum $V(\phi_-)$. 
Then, the lump solution for large values of $s$ is very close to the true vacuum at infinity and at the center of the lump. This, together with the plateau character at the center, which increases with $s$, leads that this lump solution can be aproximated  to a sequence of kink-antikink and antikink-kink pairs of the $\phi^4$ theory.
All this reasoning is in accord to a metastable lump, with a lifetime to decaying to the true vacuum that increases with $s$. Metastable lumps have already been reported as one of possible outcomes of the scatering of two-kink states \cite{bmm, mo}.


\section{Lump-lump collisions}


In this Section we will analyze lump scattering for $0.5<s<9$, covering all qualitatively different configurations described in the analysis of the Figs. \ref{fig_sol_static}-\ref{phi_V_s}. To solve the equation of motion Eq. (\ref{eom}), we use the following initial conditions:
\begin{eqnarray}
\phi_s(x,0)=\phi_{L,s}(x+x_0,v,0) + \phi_{L,s}(x-x_0,-v,0),\\
\dot{\phi}_s(x,0)=\dot{\phi}_{L,s}(x+x_0,v,0) + \dot{\phi}_{L,s}(x-x_0,-v,0),
\end{eqnarray}
where 
\begin{eqnarray}
\phi_{L,s}(x,v,t)&=&[\tanh(\gamma (x-vt)+s)-\tanh(\gamma (x-vt)-s)]/2,\\
\dot{\phi}_{L,s}(x,v,0)&=&\frac {\partial }{\partial t}\phi_{L,s}(x,v,t)_{|t=0}
\end{eqnarray}
 and $\gamma=(1-v^2)^{-1/2}$.
These conditions correspond to a pair of lumps: one centered at $ x=-x_0 $ with velocity $v$, 
and another at $ x=x_0 $ with velocity $-v$. We use a 4th order finite-difference method with a spatial step $\delta x\approx 0.05$ and we considered a $s-$dependent $x_0$ given by  $x_0(s)=14 + s$. The s-dependence of $x_0$ is justified for the following reasons: i) for small $s$, the lump solution is unstable and the scattering must occur before the lump decay completely. Then, a small $x_0$ is desirable. ii) for large $s$, the lump solution acquires a flat plateau arount its maximum that increases with $s$ (see the Fig. \ref{fig_sol_static}). Then, for large $s$ it is necessary to consider larger values of $x_0$. For time dependence we used a 6th order symplectic integrator method, being the time step $ \delta t=0.02 $. In the following we report some interesting phenomena for lump-like scattering.

\begin{figure}
	\includegraphics[width=8cm]{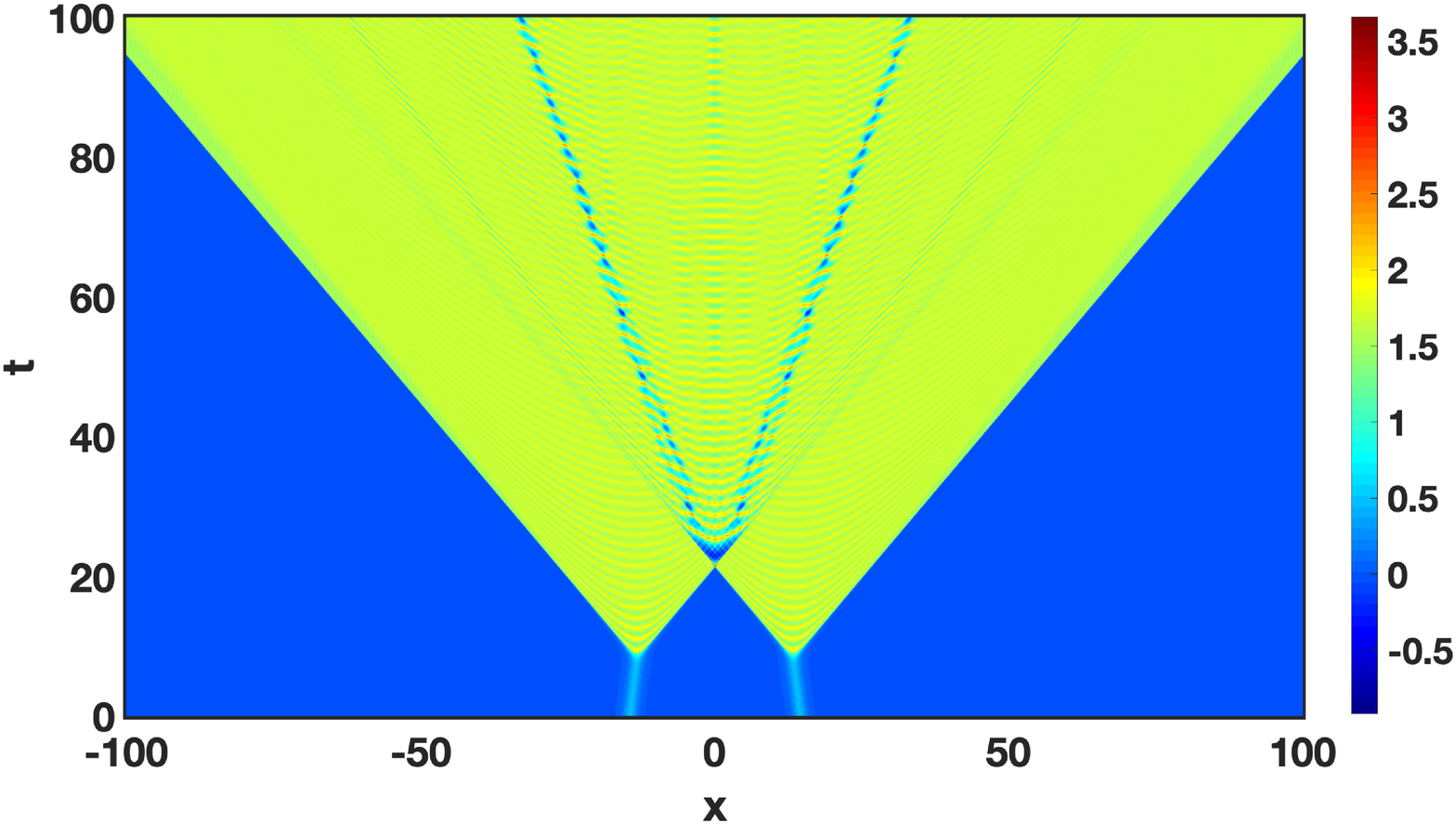} 
	\includegraphics[width=8cm]{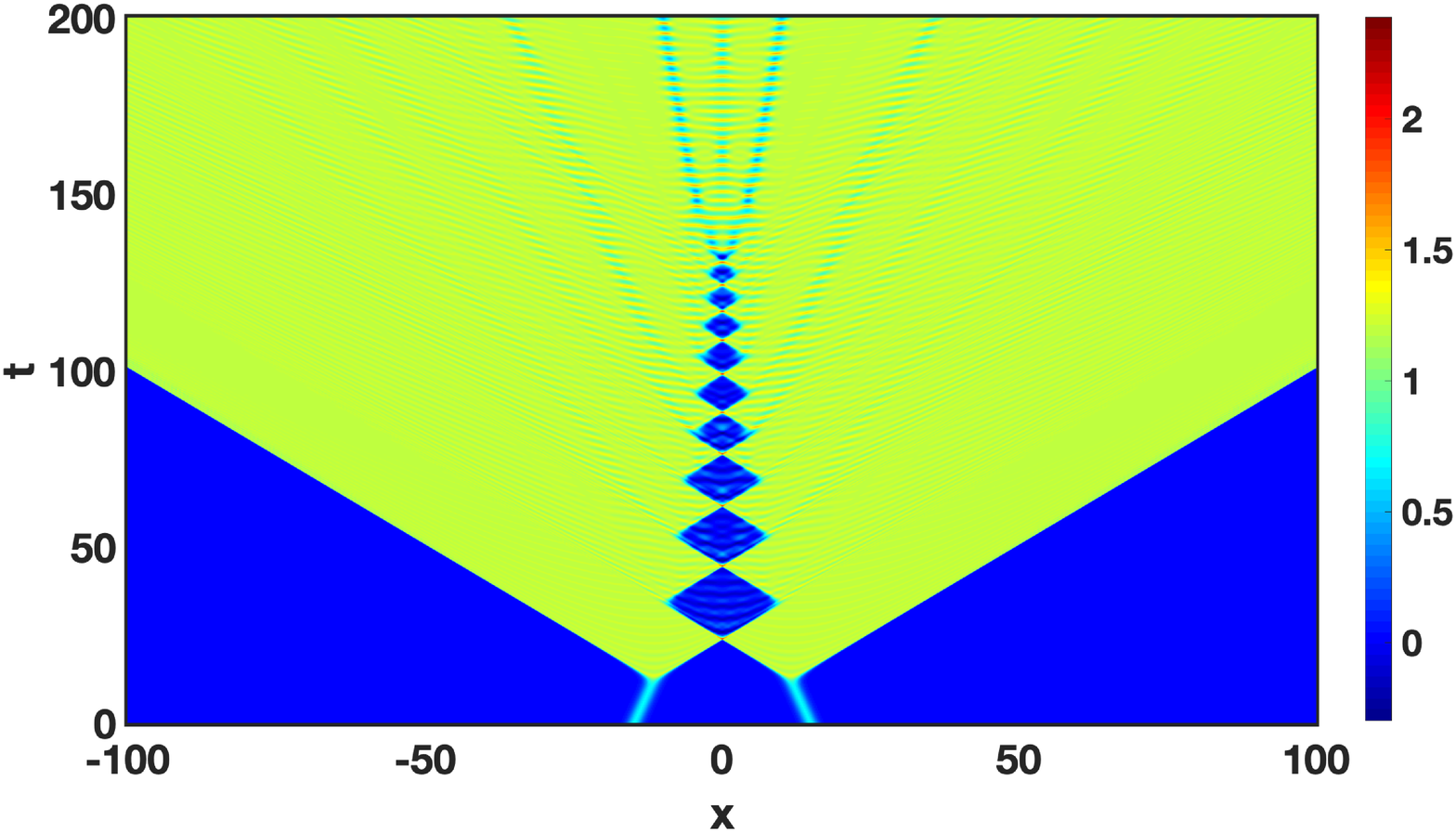} 
	\caption{Lump-lump collision with (a) $v=0.15$ for $s=0.5$ and (b) $v=0.27$ for $s=0.82$.}
	\label{small_s}
\end{figure}

For small values of the parameter $s$, the unstability of the lump can be seen in the Figs. \ref{small_s}a-b. The lump-lump pair travel towards each other for a short time and then lose their initial shape even before the collision. In addition, we can observe the formation of small oscillations with the emission of some radiation. The pattern of emission of radiation can be simple, as in the Figs. \ref{small_s}a, or more intrincate, as in the Figs. \ref{small_s}b, where at $x=0$ it appears a decaying state oscillating between $\phi_0$ and $\phi_+$ until a final collapse to $\phi=\phi_+$ followed by the production of two oscillations. 

\begin{figure}
	\includegraphics[width=8cm]{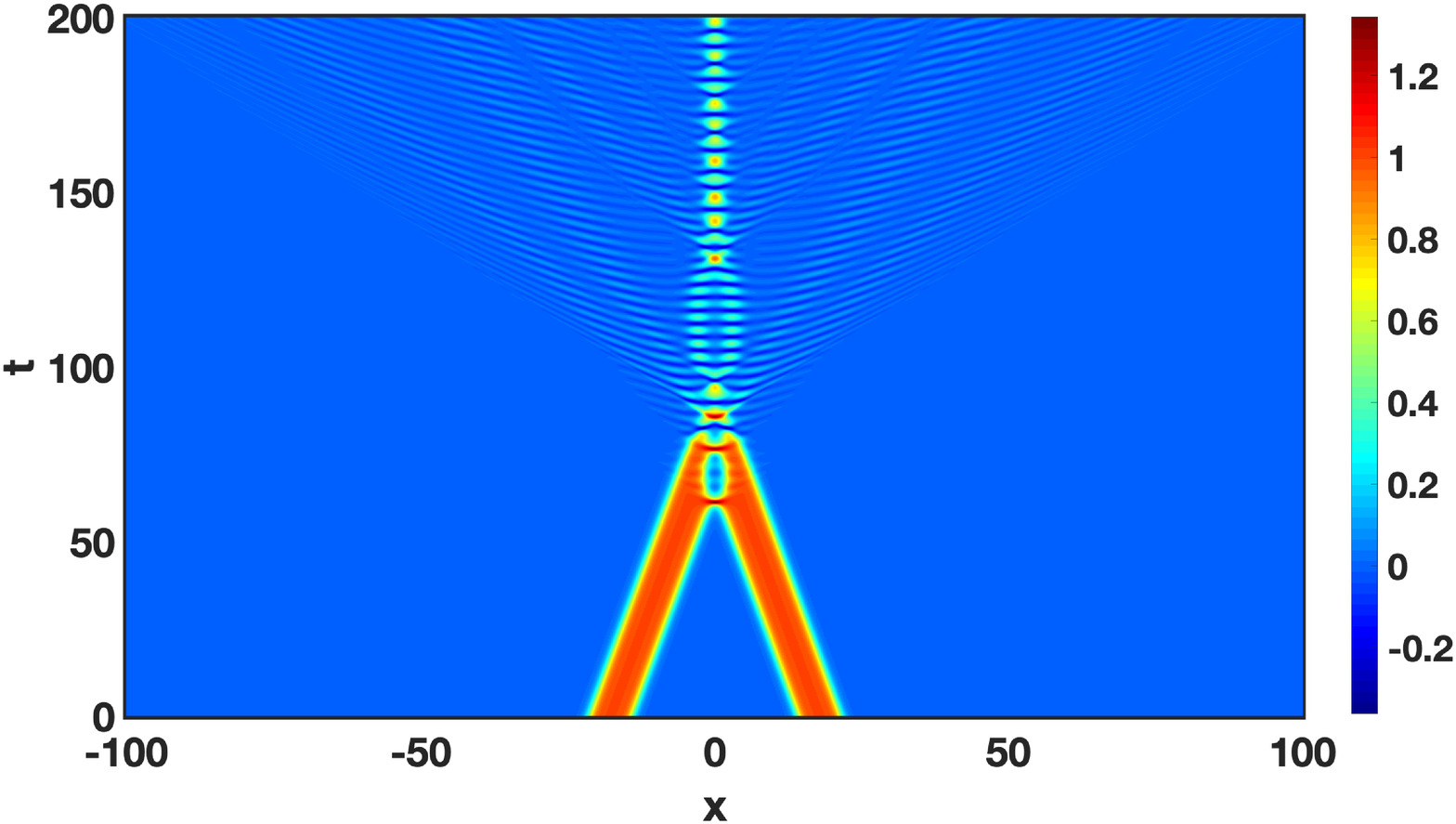}
	\includegraphics[width=8cm]{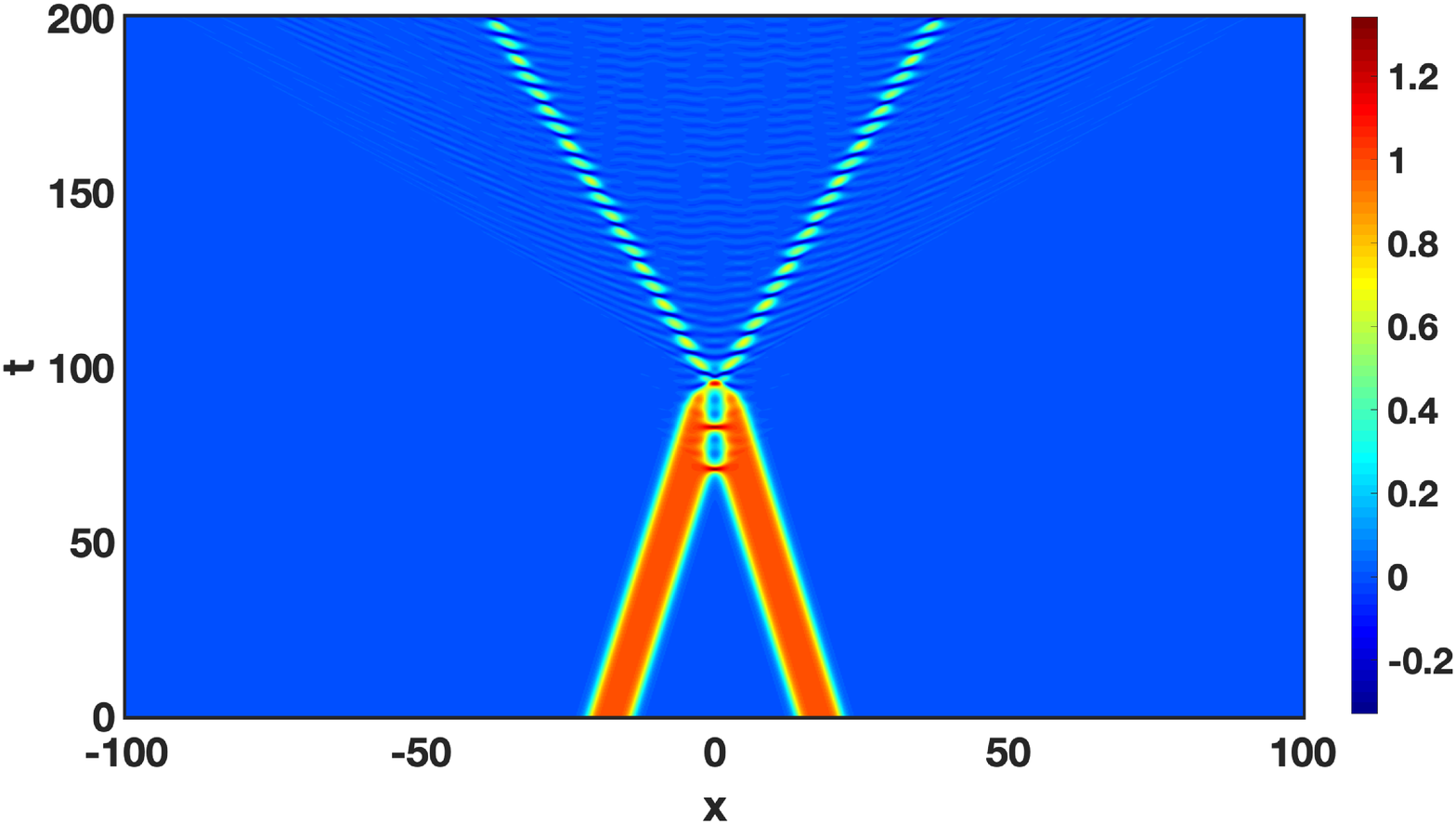}
	\includegraphics[width=8cm]{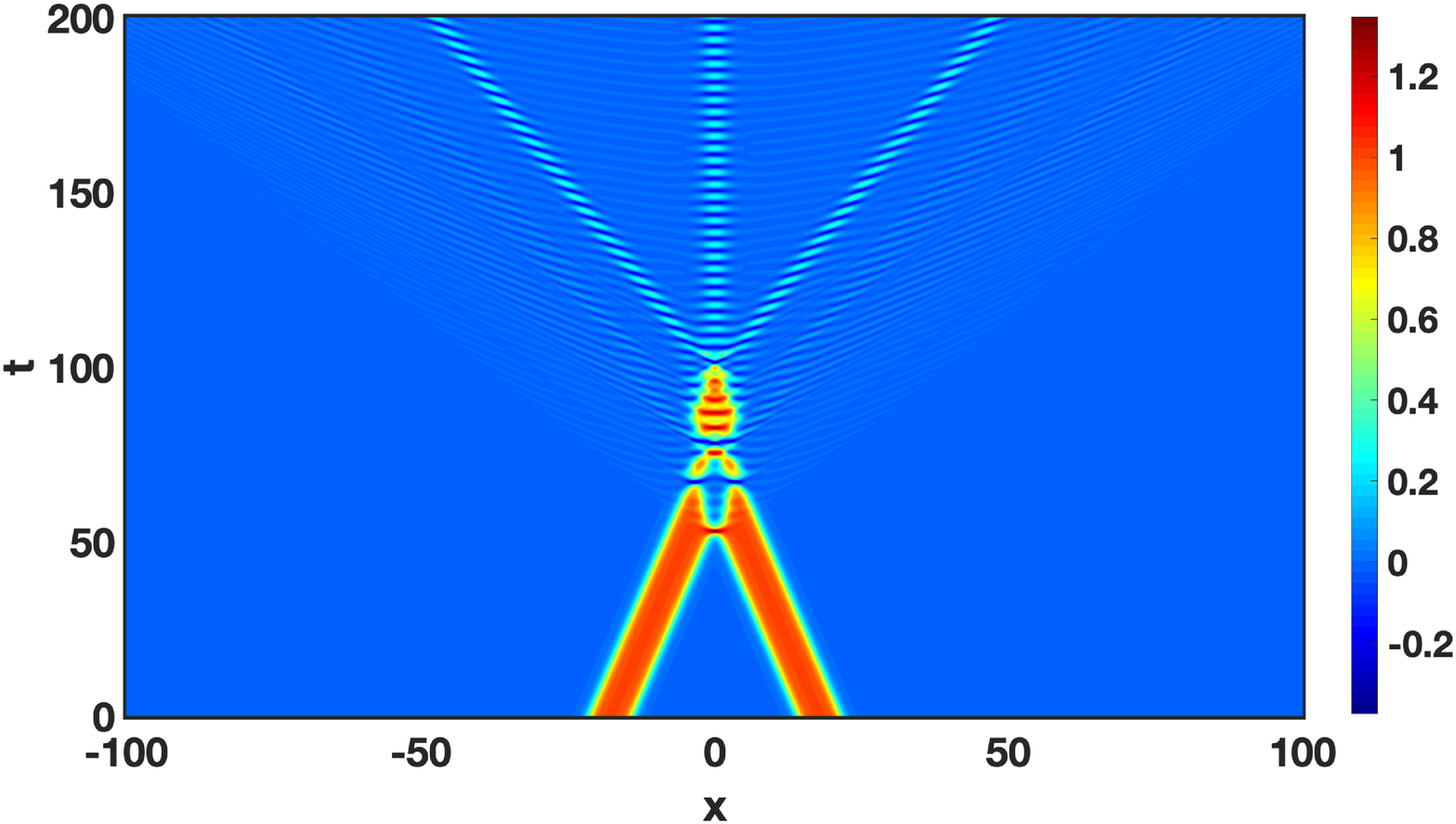}
	\caption{Lump-lump collision with (a)  $v=0.217$, producing a bion state at $x=0$, (b)  $v=0.1861$, leading to two oscillations and (c) $v=0.254$, with three oscillations. In all figures we fixed $s=3.75$.}
	\label{osc}
\end{figure}

Some examples for a large value of $s$ is presented in the Fig. \ref{osc}. There one sees that the lump solutions propagate without radiating significatively before scattering, as expected from a metastable solution. After the scattering one can see the production of a central bion (Fig. \ref{osc}a), two (Fig. \ref{osc}b) and three (Fig. \ref{osc}c) escaping oscillations along with continuously emitted radiation.  Note that the bion has a more irregular aspect than the propagating oscillations. In all cases one can see that after the scattering the output states are acompanied by a larger rate of radiation emission in comparison to the travelling lumps.

\begin{figure}
	\includegraphics[width=8cm, height=6cm]{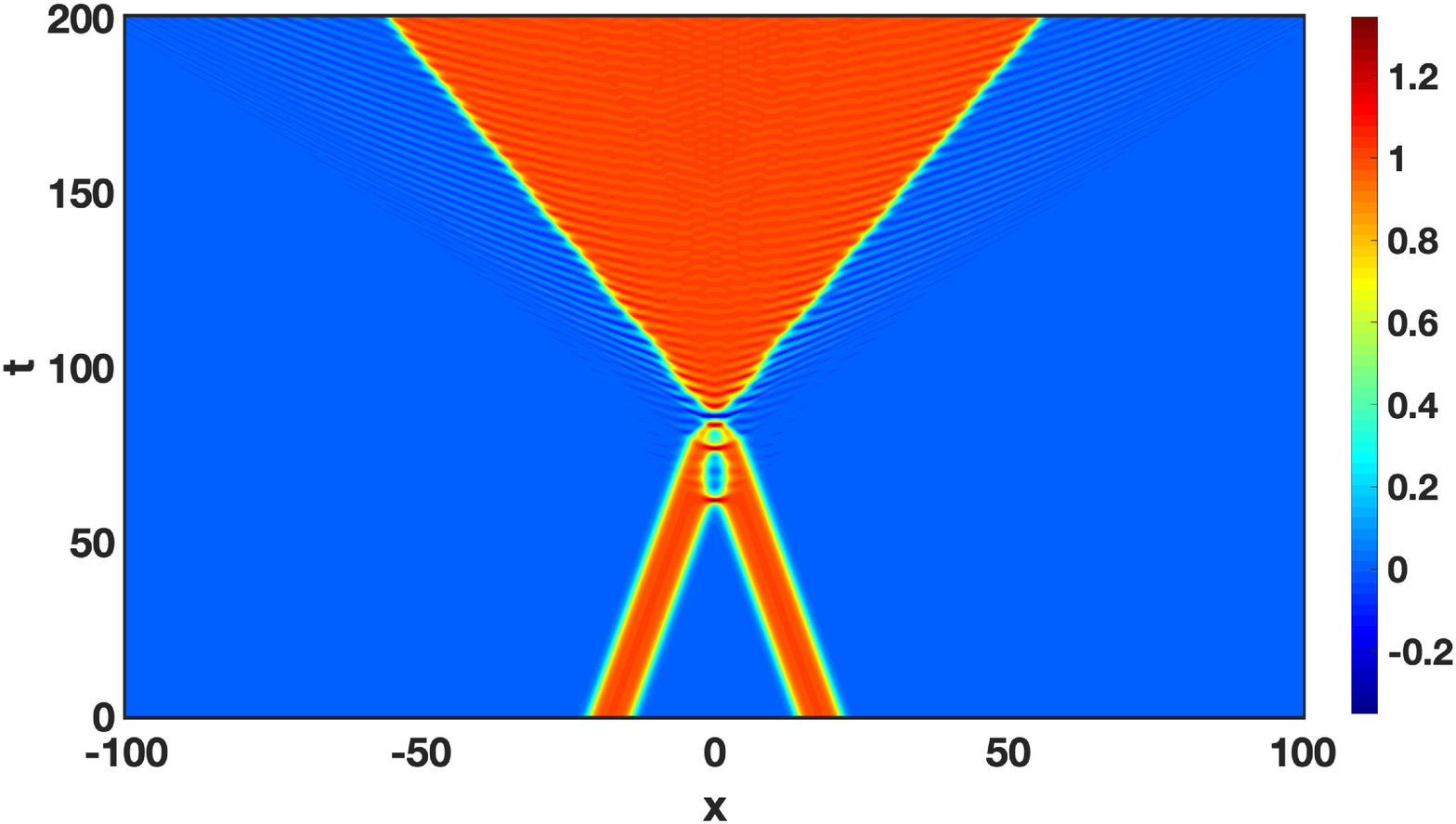}
	\includegraphics[width=8cm, height=6cm]{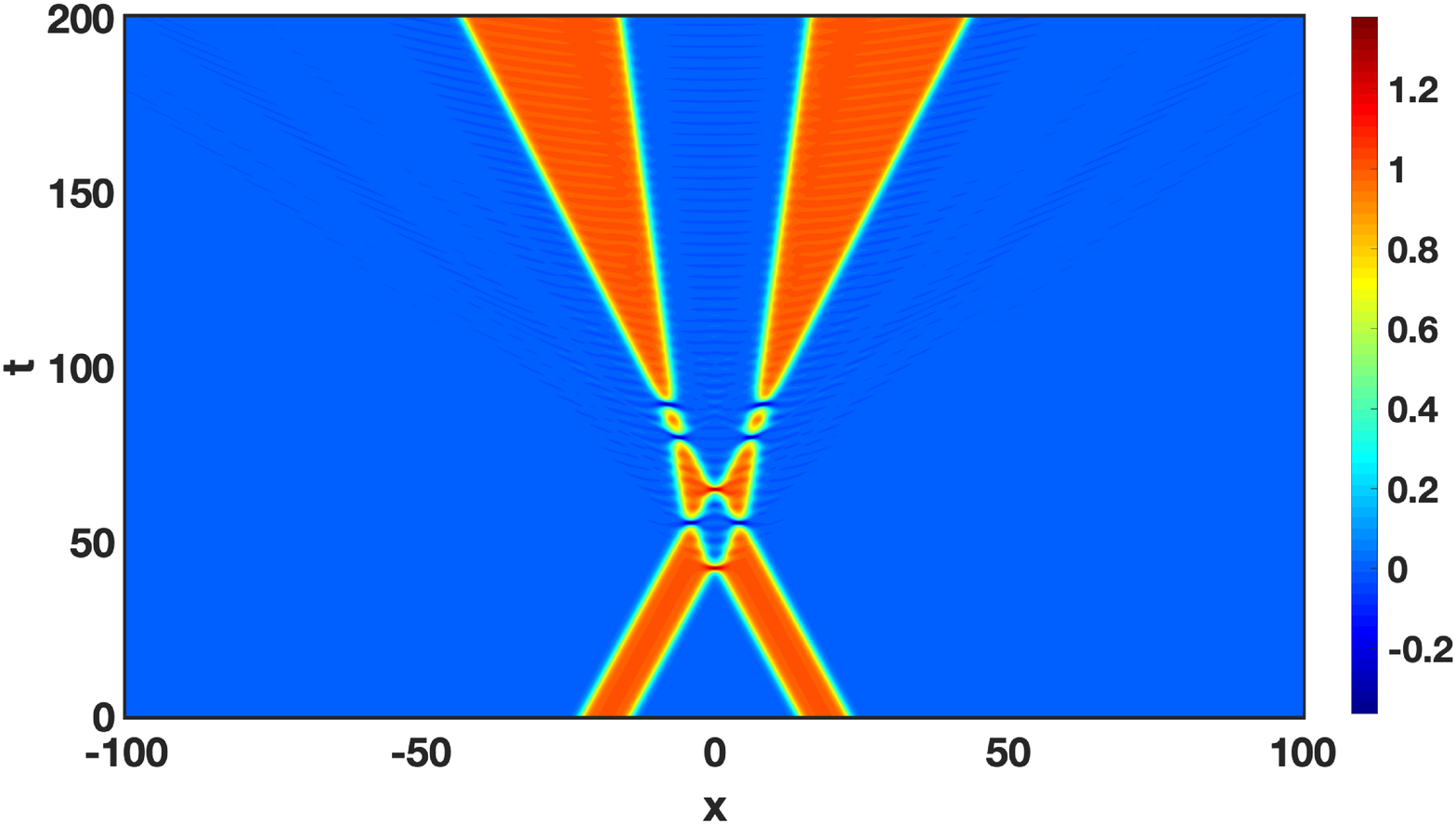}
	\caption{Lump-lump collision with (a)  $v=0.215$ for $s=3.75$  and (b)  $v=0.325$ for $s=4.5$.}
	\label{pair}
\end{figure}

For some large values of $s$ one can also see the production of one or two kink-antikink-like pairs. For instance, in the Fig. \ref{pair}a one sees that after the interaction, the lump-lump pair produces a kink-antikink-like pair that escape to infinity,  accompanied by radiation. In the Fig. \ref{pair}b we have the production of two kink-antikink-like pairs.

\begin{figure}
	\includegraphics[width=15cm]{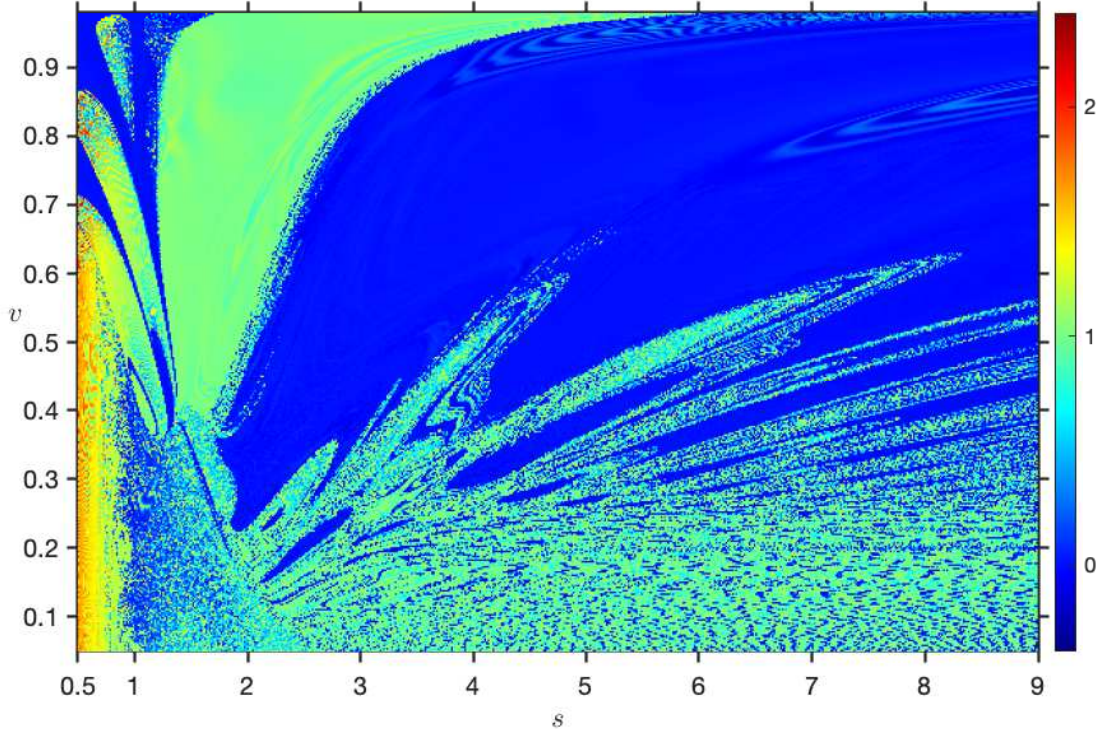}
	\caption{Phase diagram $v\times s$, showing the final state of the scalar field at $\phi(x=0,t=\frac{x_0(s)}{v} + 150)$.}
	\label{mosaico}
\end{figure}

The Fig. \ref{mosaico} is a phase diagram $v\times s$ that shows the final state of the scalar field at $x=0$.  The unstable behavior of the lump is evident for $s \lesssim 0.7$ and $v \lesssim 0.65$ and results in the final state as the true vacuum, like in the  Figs. \ref{small_s}a-b. Since the true vacuum $\phi_-$ grows for small values of $s$, as shown in the Fig. \ref{phi_V_s}a, this behavior is represented in the diagram in the regions yellow and red.  
 The blue region can indicate two possibilities: i) the production of two propagating oscillations, like in the example of the Fig. \ref{osc}b or ii) the production of two kink-antikink-like pairs - see the Fig. \ref{pair}b. The green region in the Fig. \ref{mosaico} corresponds to the occurrence of only one kink-antikink-like pair, like the Fig. \ref{pair}a. Regions with pixels randomly distributed between green and blue correspond to oscillatory patterns at $x=0$, like the Figs. \ref{osc}a (one bion state) or \ref{osc}c (the production of three oscillations). This pattern is observed in the phase diagram for $v \lesssim 0.3$ and  $1 \lesssim s \lesssim 9$. However, this behavior changes for large values of $v$. There, for small $s$, we notice the appearance of a large region in green, indicating the presence of production of one kink-antikink pair. On the other hand, the increasing of $s$ favors the appearance of two oscillations or two kink-antikink-like pairs, identified in blue.

The diagram of the Fig. \ref{mosaico} shows that metastable lumps, where $s\gtrsim 3$, has an intrincate pattern of scattering for small velocities, where the lump-lump pair has more time to interact. On the other hand, for ultralarge velocities, the diagram assumes a blue pattern, showing that it is very difficult to produce oscillations at $x=0$ or just one kink-antikink pair. In these situations, metastable lumps will produce or a pair of oscillations, or two kink-antikink pairs. For even larger values of $s$, the results tend to agree with those of the scattering of two kink-antikink pairs in the $\phi^4$ model, where for ultralarge velocities each pair of defects scatter inelastically, resulting in $\phi=0$ at $x=0$. This results in the blue region in the right upper corner of the Fig. \ref{mosaico}.
Also in the limit of large $s$, but now for low velocities, the field oscillates at $x=0$, resulting in the mixture of blue and green regions at the right lower corner of the Fig. \ref{mosaico}. For moderate velocities and $s \sim 9$, one can see some windows in velocity where the field oscillates around $x=0$ intercalated by blue regions. The presence of such fringes in the diagram is also evident in the figure for $4 \lesssim s \lesssim 9$.

\begin{figure}
	\includegraphics[width=10cm]{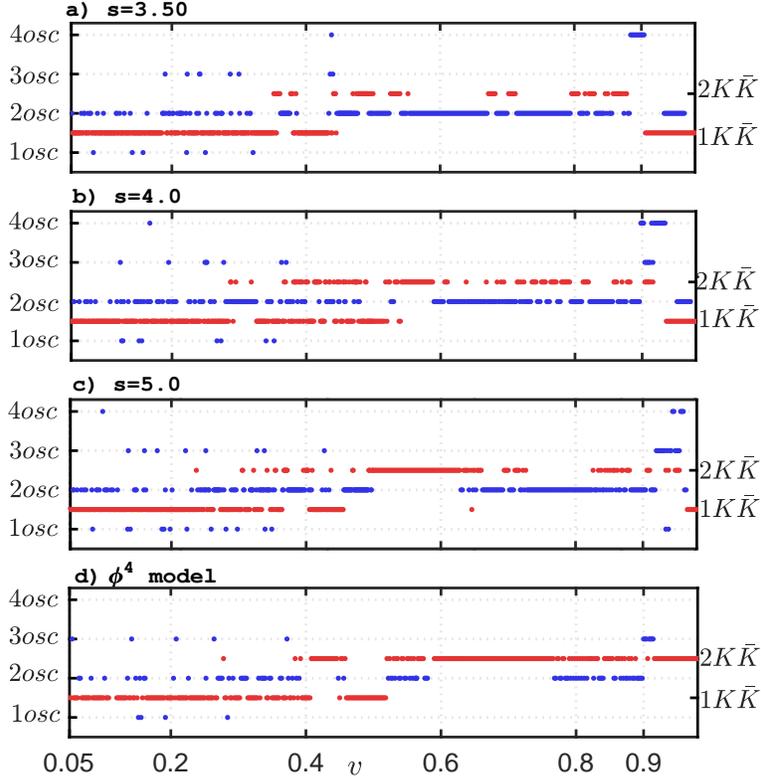} 
	\caption{Scattering of metastable lumps: the number of propagating oscillations (blue points) and the number of kink-antikink-like pairs ($K\bar K$, red points) as a function of $ v $ for a) $s=3.5$, b) $s=4.0$ and c) $s=5.0$. For comparison, we present in d) results for the scattering of two kink-antikink pairs in the $\phi^4$ model, corresponding to the limit $s\to\infty$. }
	\label{vno}
\end{figure}

In order to better characterize the regions of the diagram of the Fig. \ref{mosaico} corresponding to metastable lumps, in the Fig\ref{vno}a-c we present the number of produced propagating oscillations and kink-antikink pairs as a function of the initial velocity for large values of $s$. For comparison we present in the Fig. \ref{vno}d  the results for scattering of two kink-antikink pairs in the $\phi^4$ model, corresponding to $s\to\infty$. The Fig. \ref{vno} shows that, for low initial velocities, it is more probable the production of one kink-antikink pair intercalated for some specific velocities with the production of two oscillations for small values of $s$. This agrees with the random sector of the diagram of the Fig. \ref{mosaico}, a green region with some pixels in blue.
Also, we note the presence of windows in velocity for the production of two oscillations, corresponding to the fringes in blue in the Fig. \ref{mosaico}. For large velocities, the Fig. \ref{vno}a shows that, for $s=3.5$, it is more frequent the occurrence of two propagating oscillations than the production of two kink-antikink pairs. Both situations correspond to the blue region of the upper part of the Fig. \ref{mosaico}. Comparing the Figs. \ref{vno}a-d we see that the increasing of $s$ reduces the occurrence of two propagating oscillations, favoring the production of two kink-antikink pairs.  The production of one or three oscillations are events with lower frequency. Also events with four oscillations was registered, but they are rare, occurring only for very specific velocities.


\section{Conclusions}


In this work we have examined the effect of scattering of metastable lumps in a model depending on a parameter $s$ with a false vacuum. The parameter $s$ has a significant impact on the spectrum of small linear perturbations and in the collision process. The solutions are unstable for small $s$ ($s\lesssim2$), with a negative mode and two positive modes. For $s \gtrsim 2$ the solutions are metastable lumps, with the negative mode very close to zero. Moreover, the model for $s\to\infty$ recovers the $\phi^4$ model and the lump solution turning to a $\phi^4$ kink-antikink pair.  Our results show that for small values of $s$, the lump solutions clearly exhibit an unstable pattern, implying that when unstable lumps move towards each other, they do not retain their initial shape until the point of collision. The increase in $s$ favors long-lived lump states with two positive vibrational modes that tend to the $\phi^4$ vibrational mode frequency. In this scenario, the lump-lump collision can produce: i) an oscillating bion state at $x=0$, ii) one or two kink-antikink-like pairs, and iii) two or more (we bserved up to four) oscillations that propagate from the collision point. Our numerical results are presented in the Figs. \ref{mosaico} and \ref{vno}, which indicates that when $s$ is small, a kink-antikink-like pair is the most likely final state. Larger  values of $s$, on the other hand, favor the formation of two kink-antikink-like pairs and propagating oscillations.
Metastable lumps in this model have an intricate scattering structure and are characterized by the scalar field with the unusual aspect of a plateau around the maximum that grows with $s$. The lifetime of the produced oscillations is also large in comparison to the fast decay of the bell shaped unstable lumps, observed for small values of $s$. This means that the metastable aspect of the false vacuum for large $s$, put in evidence in the stability analysis of the lump states, can survive even after scattering, producing long-lived states.


\section{Acknowledgements}


A.R.G thanks CNPq (brazilian agency) through Grants $437923/2018-5$ and $311501/2018-4$ for financial support. This study was financed in part by the Coordena\c c\~ao de Aperfei\c coamento de Pessoal de N\'ivel Superior - Brasil (CAPES) - Finance Code 001. F.C.S. and A.R.G thank FAPEMA - Funda\c c\~ao de Amparo a Pesquisa e ao Desenvolvimento do Maranh\~ao through Grants PRONEM $01852/14$, Universal $00920/19$, $01191/16$ and $01441/18$.


\end{document}